\documentclass[aps,physrev,preprint,groupedaddress]{revtex4-2}
\usepackage{hyperref}
\usepackage{amsmath}
\usepackage{graphicx}

\begin{document}


\title{Square matrix-based six-dimensional convergence map for nonlinear beam dynamics analysis}


\author{Jinyu Wan}
\email[]{wanjinyu@ihep.ac.cn}
\affiliation{Institute of High Energy Physics, Chinese Academy of Sciences, Beijing, China, and University of the Chinese Academy of Sciences, Beijing, China}

\author{Yue Hao}
\email[]{haoy@frib.msu.edu}
\affiliation{Facility for Rare Isotope Beams, Michigan State University, East Lansing, USA}


\date{\today}

\begin{abstract}
The square matrix-based convergence map (CM) method has proven effective in characterizing nonlinear dynamics in several 4-D dynamical systems. However, when time-dependent perturbations, such as crabbing kicks in colliders, are present, a comprehensive 6-D analysis becomes essential to accurately capture the coupling between transverse and longitudinal motions. In this work, we extend the CM method to the full 6-D phase space by employing an eigen-decomposition-based formulation of the square matrix combined with iterative procedures. The proposed 6-D CM approach is first validated using a simplified crabbing map and subsequently applied to the dynamic aperture (DA) study of the future Electron-Ion Collider (EIC). The CM results show qualitative consistency with frequency analysis in identifying major resonance structures and stability boundaries by using only one-turn map, suggesting that the 6-D CM can serve as a fast indicator for nonlinear beam dynamics analysis.
\end{abstract}


\maketitle

\section{Introduction}
Understanding long-term stability in nonlinear dynamical systems \cite{robinson1998dynamical, lichtenberg2013regular} remains a critical challenge, particularly in the dynamic aperture (DA) study of particle accelerators \cite{todesco1996dynamic}. Currently, direct particle tracking remains the most reliable approach for DA evaluation, which is often computationally intensive. For lepton rings, the evaluation usually takes a few thousand of turns corresponding to the damping time. While in hadron rings, the evaluation of DA can be much heavier that takes up to $10^6$ turns \cite{chou1989dynamic, barranco2016long, crabcavityDA}. 

Over the past decades, researchers have developed faster techniques for predicting long-term stability, such as canonical perturbation theory, Lie operator methods, power series expansions, and normal form \cite{ruth1986single, guignard1978general, schoch1958theory, dragt1988lie, berz1989differential, chao2002lecture, bazzani1994normal, forest1989normal}. However, these methods often suffer from limited accuracy due to insufficient expansion orders, complications from resonance crossings, or inaccuracies at large amplitudes. In addition, the long-term evolution of DA itself has been studied in hadron rings, including inverse-logarithmic scaling models \cite{giovannozzi2012proposed} and later refinements aimed at extrapolating DA evolution beyond the turn range directly accessible by numerical tracking \cite{bazzani2019advances}.

For particles undergoing quasi-periodical motion, frequency map analysis(FMA) \cite{robin2000global, laskar2003frequency} based on Numerical Analysis of Fundamental Frequencies (NAFF) \cite{laskar1992the} has been applied on a grid of initial conditions to retrieve the shifts of the dominant frequency, as an indicator for identifying the stable region, resonances and chaotic zones \cite{comunian2001frequency, brunelle2010non, shatilov2011application}.  More recently, data-driven indicators have also been investigated, including indicators based on round-off and reversibility errors for nonlinear beam dynamics \cite{li2022data, hwang2020rapidly}, as well as broader recent overviews of chaos indicators for the CERN Large Hadron Collider (LHC) \cite{montanari2025chaos}. In parallel, machine-learning-based surrogate approaches have also been explored for predicting DA-related stability regions, including fast DA evaluation in light-source lattices \cite{wan2022machine} and reservoir-computing models for hadron rings. \cite{casanova2023ensemble}.

In recent years, notable work has been made in the application of the square-matrix method \cite{yu2017analysis} for the efficient analysis of periodic nonlinear dynamical systems, particularly in the identification of high-order resonances. Based on this foundation, the convergence map (CM) method was developed \cite{yu2023convergence}. The CM method employs an iterative transformation that converts nonlinear trajectories into a nearly pure rotational motion, with the residual convergence error serving as an indicator of stability. This framework has been successfully validated in several 4-D dynamical systems, including the 4-D Henon map \cite{anderson2022study} and the transverse dynamics of the NSLS-II storage ring \cite{yu2019theory, yu2022progress}. Compared with conventional particle tracking and FMA techniques, the CM method offers much faster computational efficiency as it only requires one-turn transfer map in the calculation. 

In modern ring colliders, synchro-betatron coupling between transverse and longitudinal dynamics arises from various effects such as accelerating cavities located at dispersive location, beam-beam interactions and other transverse fields with arriving time dependence \cite{barranco2016long}. In recent collider designs, the hadron beam is designed to collide with another hadron beam or electron beam, such as the LHC high luminosity upgrade and the Electron-Ion Collider (EIC) \cite{berg2023lattice} project. Crab cavities, which provide a time-dependent dipole field, are used to compensate for the loss of geometric luminosity due to the crossing angle. Previous studies show that the high harmonic magnetic moment in the crab cavity will affect the DA \cite{crabcavityDA}, and might generate additional resonance lines affecting long-term stability of the beam \cite{xu2021study}. Using a reduced 4-D analysis can be insufficient to capture such effects. 

In this work, we extend the CM from 4-D to the full 6-D phase space to study these synchro-betatron coupling effects. The proposed 6-D CM formulation is based on eigen-decomposition of the one-turn transfer map, enabling more accurate reconstruction of an upper triangular square matrix compared with the Courant-Snyder form in the original 4-D CM method. By combining this formulation with iterative convergence analysis, we reconstruct approximate action-angle variables that provide local coordinates associated with approximately quasi-periodic motion in the coupled 6-D space. This extension allows the CM to identify subtle high-order resonances and nonlinear couplings that cannot be resolved through conventional FMA. The approach is first validated using a simplified crabbing map containing time-dependent multipole components and is subsequently applied to the hadron storage ring of the EIC to evaluate its DA and the influence of nonlinear crab cavity fields.

The content is organized as follows. A detailed description of the 6-D CM method will be given in Section \ref{sec:method}. The method will then be implemented on a simplified 6-D nonlinear map, abstracted from crab cavity kicks, in Section \ref{sec:CM_on_Map}. In Section \ref{sec:CM_on_HSR}, this method is also applied to the hadron storage ring of the EIC to study the impact of multipole components in crab cavities and evaluate its DA.

\section{Method\label{sec:method}}
\subsection{Approximated action-angle variables using square matrix}
For periodic dynamical systems, such as synchrotron accelerators, the one-turn map can be expressed as a set of polynomial equations and rewritten as a square matrix. For a dynamical system described by the canonical variable $x$ and its momentum $p$, if we use the complex Courant-Snyder variable $z=\frac{x}{\sqrt{\beta}}-\frac{i}{\sqrt{\beta}}(\alpha x+p)$ and its conjugate $z^*$ to replace $x$ and $p$, $\{z, z^*,z^2,zz^*,z^{*2},z^3...\}$ will form a new column $Z$ where the one-turn map can be described by a square matrix $M$,
\begin{equation}
    Z=MZ_0.
\end{equation}

In principle, the polynomial square-matrix representation is infinite dimensional. In practical calculations, it is truncated at a finite polynomial order and therefore provides only a local finite-order representation of the nonlinear one-turn map. In the present work, we do not attempt to determine the spectrum of the exact Koopman operator, nor do we interpret the eigenvectors of the truncated square matrix as globally convergent Koopman eigenfunctions. Instead, these eigenvectors are used as a computational basis to construct nonlinear coordinates in which the local dynamics is closer to a pure rotation. The subsequent convergence procedure is then used to quantify the residual deviation from this rotational behavior.

All square matrices can be transformed into Jordan form with a transformation matrix $U$, an eigenvalue $\mu$, and a Jordan matrix $\tau$ \cite{kaagstrom1980algorithm}
\begin{equation}
    UM=e^{i\mu I+\tau}U,
\end{equation}
where $I$ is the identity matrix and the Jordan matrix has the form
\begin{equation}
    \tau= \begin{bmatrix}
0 & 1 & 0 & ... & 0\\
0 & 0 & 1 & ... & 0\\
0 & 0 & ... & ... & 0\\
0 & 0 & 0 & ... & 1\\
0 & 0 & 0 & ... & 0
\end{bmatrix}.
\end{equation}
Then Eq. 1 gives
\begin{equation}
    UZ=UMZ_0=e^{i\mu I+\tau}UZ_0.
\end{equation}
A transformation is defined as
\begin{equation}
    W\equiv UZ,
\end{equation}
where $W$ represents the projection of $Z$ onto the invariant subspace spanned by the left eigenvectors $u_j$ given by the rows of the matrix $U$. Therefore, each row of $W$ is represented as $w_j=u_jZ$, which is a polynomial of $\{z, z*,...\}$.

The new vector after $n$ turn becomes
\begin{equation}
    W=e^{n(i\mu I+\tau)}W_0.
\end{equation}

According to KAM theory, for sufficiently small perturbations and under appropriate non-resonance and non-degeneracy conditions, a set of invariant tori survive under perturbation and supports quasiperiodic motion \cite{broer2004kam}. In that regime, the transformed variables are expected to remain close to a pure rotation, so that $|w_j|$ is approximately invariant and the phase advance differs from the linear value by a small correction $\mu+\phi$, where $\phi\ll \mu$ is the correction value. Therefore, after $n$ turns, approximately, the vector $W$ only changes by a phase factor of $e^{in(\mu+\phi)}$,
\begin{equation}
    W=e^{n(i\mu I+\tau)}W_0\cong e^{in(\mu+\phi)}W_0.
\end{equation}
From Eq. 7, we have
\begin{equation}
    \tau W_0\cong i\phi W_0.
\end{equation}
Using the property of the Jordan matrix in Eq. 3, the above equation can be explicitly written as
\begin{equation}
    \tau \begin{bmatrix}
w_0 \\
w_1 \\
... \\
w_{m-1}  
\end{bmatrix} = \begin{bmatrix}
w_1 \\
w_2 \\
... \\
0  
\end{bmatrix} \cong \begin{bmatrix}
i\phi w_0 \\
i\phi w_1 \\
... \\
i\phi w_{m-1}  
\end{bmatrix},
\end{equation}
where $w_j$ represents the $j$ th row of $W_0$. $i\phi$ is then derived by comparing the two sides,
\begin{equation}
    i\phi\cong\frac{w_1}{w_0}\cong\frac{w_2}{w_1}\cong\frac{w_3}{w_2}\cong...
\end{equation}

For a 6-D scenario, Eq. (7) is derived for the planes $x$, $y$, and $z$, resulting in three sets of polynomials {$w_{x0}, w_{x1},...$}, {$w_{y0}, w_{y1},...$} and {$w_{z0}, w_{z1},...$}. These polynomials can be used as a set of approximated action-angle variables. Their trajectories form a circle with a small deviation from a pure rotation. It was reported in \cite{yu2023convergence} that near the boundary of the stable region, the deviation of these actions from a pure rotation provides an important indicator of the destruction of invariant tori, which measures the stability of the trajectories.

\subsection{Square matrix with eigen-decomposition}
In realistic dynamical systems with strong nonlinearity and coupling terms, simply using Courant-Snyder variables in Eq. (1) does not necessarily result in an upper triangular square matrix $M$. To improve the precision of square matrix construction, we use a numerical decomposition of the transfer matrix $M_0$, the eigen-decomposition, to refactor $M_0$ into a canonical form. The original transfer matrix is defined as
\begin{equation}
    X=M_0X_0,
\end{equation}
where $X_0$ represents the 6-D coordinate vector with respect to the closed orbit. With eigen-decomposition, $M_0$ can be refactored as
\begin{equation}
    M_0=V\Lambda V^{-1},
\end{equation}
where $\Lambda$ is a diagonal matrix containing eigenvalues and $V$ is a complex $6\times6$ matrix. 

Similarly, the new column $Z$ can be written as $Z=V^{-1}X$. The square matrix in Section II.A is now replaced by the transfer map of the eigen-coordinates $Z$. Similar to Eq. (5), the $W$ variable is defined as
\begin{equation}
    W\equiv VZ.
\end{equation}
Here $W$ represents the same physical quantities as $X$, but expressed on the basis $Z$ through the eigenvectors $v_j$, namely,
\begin{equation}
    w_j=v_jZ.
\end{equation}

\subsection{The convergence map method}
In the previous section, a transformation $V$ is generated by eigen-decomposition to create a set of new variables $w$, so that the transfer map $\Lambda=V^{-1}M_xV$ is approximately a pure rotation and $w$ is approximated action-angle variables. Considering that the new set of $w$ variables is not unique, for a 6-D dynamical system, we use $w_x$, $w_y$ and $w_z$ to denote the approximated action-angle variables chosen. Their phases define the corresponding angle variables,
\begin{equation}
    w_j\equiv e^{i\theta_j}, j=1,2,3,
\end{equation}
where $j=1,2,3$ corresponding to the $x$, $y$, and $z$ planes, respectively.

Note that, unlike the standard definition in a Hamiltonian system, these approximated action-angle variables are a loose term, with their absolute values as actions and their phases as angles. This can be further loosely extended to use Re$(\theta)$ as angle and Im$(\theta)$ \cite{yu2023convergence}.

For a given initial condition, we seek a torus representation in which the motion is as close as possible to a pure rotation. To do so, we introduce a set of torus-angle coordinates,
\begin{equation}
    \boldsymbol{\alpha}=(\alpha_1,\alpha_2,\alpha_3)\in [0,2\pi)^3,
\end{equation}
where $\alpha_1$, $\alpha_2$, and $\alpha_3$ are sampled on a discrete three-dimensional grid. We then write the transformed motion as
\begin{equation}
    \theta_j^{(m)}=\theta_{j,ini}+\alpha_j+\delta\theta_j^{(m)}(\boldsymbol{\alpha}),
\end{equation}
where $\theta_{j,ini}$ is determined from the initial condition of the tracking, $m$ is the iteration index, and $\delta\theta_j^{(m)}$ is the correction field at iteration $m$. The initial torus is taken as,
\begin{equation}
    \delta\theta_j^{(0)}(\boldsymbol{\alpha})=0.
\end{equation}

At each iteration, the current torus is mapped one turn forward. More specifically, for every grid point $\boldsymbol{\alpha}$, we first reconstruct
\begin{equation}
    w_j^{m}(\boldsymbol{\alpha})=\exp(i\theta_j^{(m)}(\boldsymbol{\alpha})),
\end{equation}
then transform from $w$-space back to the original phase-space variables, apply the nonlinear one-turn map, and transform the result again into the $w$-variables. This gives the one-turn image $w_{j,next}^{(m)}(\boldsymbol{\alpha})$.

The local phase advance of the current torus is then measured by
\begin{equation}
    \phi_j^{(m)}(\boldsymbol{\alpha})=-i\log(\frac{w_{j,next}^{(m)}(\boldsymbol{\alpha})}{w_{j}^{(m)}(\boldsymbol{\alpha})}), j=1,2,3.
\end{equation}

If the torus were pure rotation, $\phi_j^{(m)}$ would be independent of $\boldsymbol{\alpha}$. We therefore decompose it into a constant part and a nonuniform remainder, 
\begin{equation}
    \phi_j^{(m)}(\boldsymbol{\alpha})=w_j^{(m)}+\Delta\phi_j^{(m)}(\boldsymbol{\alpha}),
\end{equation}
where $\Delta\phi_j^{(m)}(\boldsymbol{\alpha})$ measures the deviation from a pure rotation. Numerically, $w_j^{(m)}$ is obtained from
the zero Fourier mode of $\phi_j^{(m)}$, while the nonzero Fourier components are used to update the correction field $\delta\theta_j^{(m)}$. Specifically, writing
\begin{equation}
\Delta\phi_j^{(m)}(\boldsymbol{\alpha})
=
\sum_{\boldsymbol{k}\neq\boldsymbol{0}}
\widetilde{\Delta\phi}_{j,\boldsymbol{k}}^{(m)}
e^{i\boldsymbol{k}\cdot\boldsymbol{\alpha}},
\end{equation}
where $\boldsymbol{k}=(k_1,k_2,k_3)$, the corresponding Fourier correction contains the factor
\begin{equation}
\widetilde{\delta\theta}_{j,\boldsymbol{k}}^{(m)}
\propto
\frac{
\widetilde{\Delta\phi}_{j,\boldsymbol{k}}^{(m)}
}{
\exp\left[
i\boldsymbol{k}\cdot\boldsymbol{w}^{(m)}
\right]-1
},
\qquad
\boldsymbol{k}\neq\boldsymbol{0}.
\label{eq:resonance_denominator}
\end{equation}
Here
$\boldsymbol{w}^{(m)}
=(w_1^{(m)},w_2^{(m)},w_3^{(m)})$
denotes the three rotation numbers. The denominator in Eq.~(\ref{eq:resonance_denominator}) becomes small when $\boldsymbol{k}\cdot\boldsymbol{w}\simeq2\pi p$, with $p$ an integer, corresponding to the resonance condition
\begin{equation}
k_1\nu_x+k_2\nu_y+k_3\nu_z\simeq p.
\end{equation}
Therefore, resonance crossings naturally enhance the corresponding Fourier corrections and can lead to poor convergence of the iteration. In this way, resonance sensitivity enters explicitly in the recursive Fourier procedure. The correction field is iteratively updated so that the transformed motion becomes closer and closer to a pure rotation.

The convergence of the iteration is monitored by the RMS difference between two successive corrections fields,
\begin{equation}
    r^{(m)}=[\frac{1}{N_{\alpha_1} N_{\alpha_2} N_{\alpha_3}}\sum^{3}_{j=1}\sum_\alpha |\delta\theta_j^{(m)}(\boldsymbol{\alpha})-\delta\theta_j^{(m-1)}(\boldsymbol{\alpha})|^2]^{1/2},
\end{equation}
where $N_{\alpha_1}$, $N_{\alpha_2}$ and $N_{\alpha_3}$ are the numbers of grid points in the three directions. The convergence metric is defined as
\begin{equation}
    C=\min\space r^{(m)}, 1\leq m\leq M,
\end{equation}
where $M$ is the maximum number of iterations.

Small values of $C$ indicate good convergence of the torus-construction procedure and therefore suggest the existence of a nearby smooth quasi-periodic torus. Large values of $C$ indicate poor convergence, which typically occurs near resonance lines or outside the regular stable region. In this sense the convergence residual provides a practical numerical indicator of local stability.

The accuracy of this construction is necessarily limited by the polynomial truncation. Increasing the truncation order improves the formal representation of the local one-turn map, but does not necessarily improve the CM iteration, since higher-order terms also increase the dimensionality of the square matrix and numerical sensitivity of the transformed coordinates. The CM should therefore be regarded as a finite-order local stability indicator rather than a convergent reconstruction of global invariant tori. In this work, third-order maps are used because they provide a practical balance between nonlinear information and computational cost.

In summary, the CM procedure consists of three steps. First, a finite-order polynomial representation of the nonlinear one-turn map is constructed and expressed as a square matrix. Second, selected eigenvectors of this matrix define nonlinear coordinates in which regular trajectories are approximately rotational. Third, these coordinates are iteratively corrected using Fourier components of the nonuniform phase advance. The minimum residual during this iteration defines the convergence metric.

\section{Application on a 6-D toy map \label{sec:CM_on_Map}}
Crab cavities are specialized RF cavities designed to restore head-on collisions in colliders for high luminosities. Here, we investigate a 6-D periodic system that includes a crab cavity kick. A time-dependent sextupole kick is also incorporated into the crab cavity model as the multipole components in the cavity. The Hamiltonian and transfer map associated with the crab cavity kick can be expressed as
\begin{equation}
    H_{kick}(x,y,z)=\frac{\tan(\theta_{cc})}{k_c\sqrt{\beta_{cc}\beta_{IP}}}x\sin(k_cz)-b_3(\frac{x^3}{3}-xy^2)\sin(k_cz)
\end{equation}
\begin{equation}
    \delta p_x=-\frac{\tan(\theta_{cc})\sin(k_cz)}{k_c\sqrt{\beta_{cc}\beta_{IP}}}+b_3(x^2-y^2)\sin(k_cz)
\end{equation}
\begin{equation}
    \delta p_y=-2b_3xy\sin(k_cz)
\end{equation}
\begin{equation}
    \delta p_z=-\frac{x\tan(\theta_{cc})\cos(k_cz)}{\sqrt{\beta_{cc}\beta_{IP}}}+\frac{b_3k_c}{3}(x^3-3xy^2)\cos(k_cz),
\end{equation}
where $\theta_{cc}$ is the crossing angle, $k_c$ is the wave number, $b_3$ is the strength of the time-dependent sextupole kick in the crab cavity, and $\beta_{cc}$ and $\beta_{IP}$ are beta functions at the crab cavity and the interaction point (IP), respectively. 

The rest of the periodic map is made up of pure rotations with tunes of $[\nu_x,\nu_y,\nu_z]=[0.26,0.23,0.005]$. These tune values satisfy the high-order resonance condition $2\nu_x+2\nu_y+4\nu_z=1$, indicating potential resonance behaviors in the system. It is worth noting that this toy model and its tune values are introduced to test the CM method and are not intended to reproduce an operating point of any real-world accelerators. To make the synchro-betatron coupling strong enough, the strength of the sextupole term is set to 50,000 m$^{-3}$ that is an unrealistically high value.

Although the crab-cavity Hamiltonian contains trigonometric functions of the longitudinal coordinate, the square matrix is constructed from the finite-order Taylor representation of the complete one-turn map about the reference orbit. JuTrack automatically generates this high-order polynomial map and the trigonometric terms such as $\sin(k_cz)$ and $\cos(k_cz)$ are therefore expanded consistently to the same prescribed polynomial order before the square-matrix representation is constructed.

To investigate the 6-D beam dynamics of the system, especially the synchro-betatron couplings, we generate a uniform grid of $100\times100$ particles in the $x$-$z$ plane. The initial coordinates $y$ of all particles are fixed at 0.5 mm, while the initial momenta $p_x$, $p_y$, and $p_z$ are zero. The high-order transfer map for this system is constructed with JuTrack \cite{wan2025jutrack}, a novel auto-differentiable simulation code developed in the Julia programming language, and then evaluated at the chosen initial conditions. The resulting dynamics are then analyzed using the CM up to the third order. For comparison, the FMA method is also applied to this toy system. Particle trajectories of 2,000 turns obtained from particle tracking simulation are used to calculate the frequency diffusion.

\begin{figure}[!htb]
\centering
\includegraphics[width=1\columnwidth]{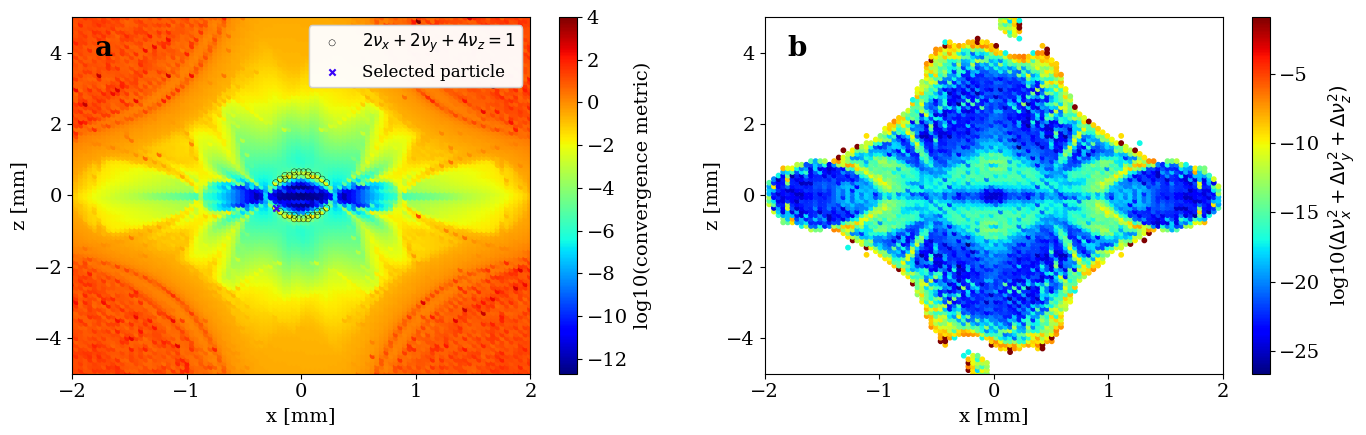}
\caption{CM and FMA results for the toy system in the $x$-$z$ plane. The left plot shows the convergence metric $C$. A large value of the convergence metric indicates divergence of the convergence map. The color scale in the right plot represents tune diffusion of the trajectories.}\label{fig1}
\end{figure}

\begin{figure}[!htb]
\centering
\includegraphics[width=1\columnwidth]{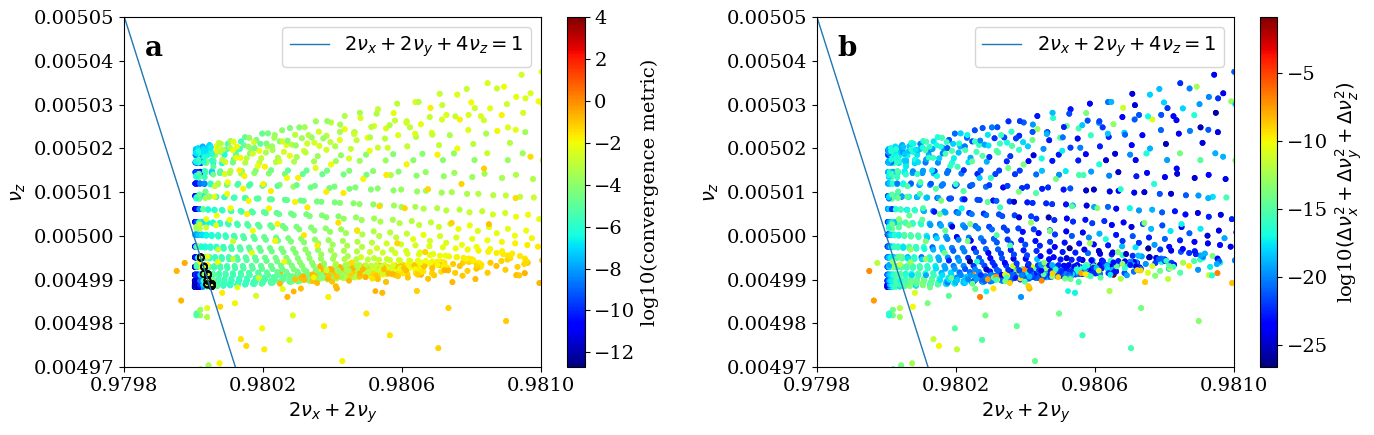}
\caption{Results of CM (left) and FMA (right) in the frequency space. The black circles in the left plot correspond to the black circles in spatial space shown in Fig. 1(a). The x-axis represents $2\nu_x+2\nu_y$ and the y-axis represents $\nu_z$.}\label{fig2}
\end{figure}

Figure 1 presents the results of applying the CM method to the toy model and its comparison with the FMA method. The results show that CM successfully detects resonances identified by FMA, validating its accuracy and effectiveness in characterizing 6-D nonlinear dynamics for this system as a qualitative alternative method. In particular, the CM method also reveals additional resonance structures not detected by FMA, such as the selected points highlighted in Fig. 1. These structures are located in regions where FMA indicates small tune diffusion.

Further insight is provided by the tune footprint shown in Fig. 2, where the highlighted points are found to cross the eighth-order resonance line $2\nu_x+2\nu_y+4\nu_z=1$. This resonance is intentionally introduced into the model to test both methods. The CM method successfully identifies this eighth-order resonance. This result suggests a high sensitivity of the CM method to resonances, especially for high-order or subtle resonances.

The purpose of the CM method is not to replace FMA or long-term tracking. Rather, it provides a complementary indicator that extracts information on resonance structures and local stability directly from the one-turn nonlinear map, without requiring long-term tracking for every initial condition.

\section{Application on the Electron-Ion Collider \label{sec:CM_on_HSR}}
The design of the Electron-Ion Collider (EIC) aims to enable collisions between polarized electrons and a wide range of ion species, including spin-polarized protons at energies up to 275 GeV. The Hadron Storage Ring (HSR) of the EIC is designed to reuse much of the existing infrastructure from the Relativistic Heavy Ion Collider (RHIC) \cite{harrison2002rhic}, with modifications to accommodate the requirements of the EIC physics program. In particular, crab cavities are employed in the EIC to restore head-on collisions for higher luminosity.

Evaluating the DA of the HSR requires long-term particle tracking, typically over $10^6$ turns, making it computationally expensive. To efficiently evaluate the DA and study the nonlinear beam dynamics of the HSR, the CM method is applied to the HSR lattice to analyze local stability across different slices in the phase space. The HSR lattice is constructed using JuTrack. The model includes the lattice nonlinearities and the selected RF cavities described below, but does not attempt to include all effects relevant to the final DA of the machine, such as a complete treatment of beam-beam interaction and all possible error sources. There are two pairs of crab cavities in the HSR lattice. For simplicity, we consider only a single pair of 197 MHz cavities in this study. Following the study of multipole components in the EIC crab cavities \cite{wu2021eic, wu2022eic}, similar to Eq. (19-21), time-dependent sextupole components with geometric strength of $10^{-3}$ m$^{-3}$ are added to the crab cavities. Only the 591 MHz cavity and five 197 MHz cavities are activated to generate longitudinal oscillations. 

\begin{figure}[!htb]
\centering
\includegraphics[width=1\columnwidth]{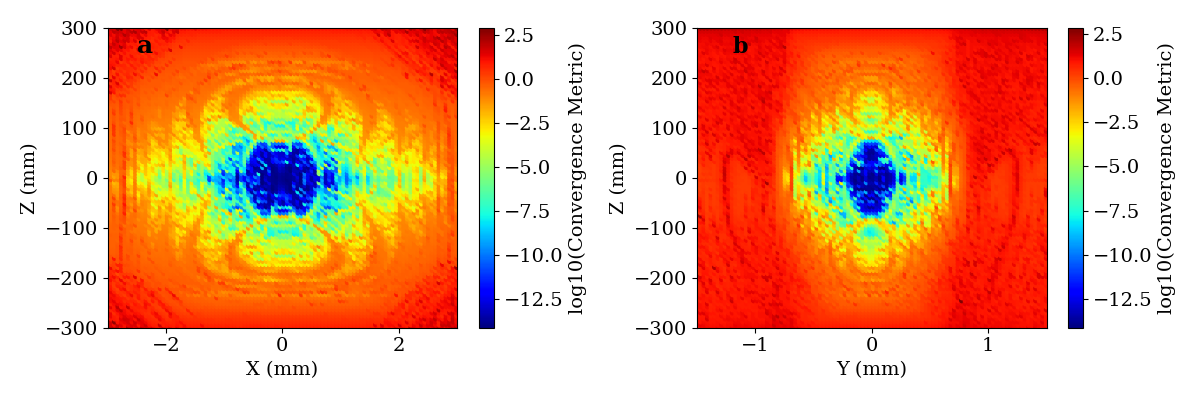}
\caption{CM analysis results for the EIC HSR in the $x$-$z$ (left) and $y$-$z$ (right) plane. The color scale from blue to red represents the convergence error to a pure rotation increasing from small to large.}\label{fig3}
\end{figure}

\begin{figure}[!htb]
\centering
\includegraphics[width=1\columnwidth]{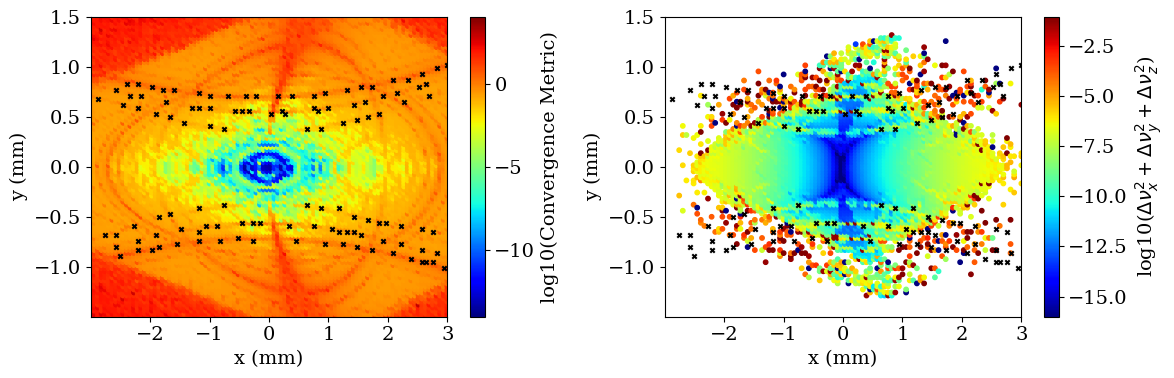}
\caption{Results of applying CM (left) and FMA (right) to the EIC HSR in the $x$-$y$ plane. The initial $z$ for all particles are fixed at 6 cm. The black points represent some selected trajectories crossing resonance lines.}\label{fig4}
\end{figure}

The CMs are calculated across different slices in the phase space. Specifically, we consider three slices in the $x$-$y$, $x$-$z$, and $y$-$z$ planes, fixing the third coordinate at an RMS beam size ($\sigma_z$, $\sigma_y$, and $\sigma_x$, respectively). The scanning ranges are [-3, 3] mm for $x$, [-1.5, 1.5] mm for $y$, and [-30, 30] cm for $z$, covering a wide region in the space compared to the DA size. The initial momenta $p_x$, $p_y$ and $p_z$ are set to 0.

Figure 3 shows the CM evaluated in two longitudinal-transverse slices of the HSR phase space, $x$-$z$, and $y$-$z$. Dark blue regions correspond to strongly convergent solutions, indicating the behavior consistent with a nearby quasi-periodic torus, whereas yellow-red regions represent loss of convergence and are associated with resonance crossing or very large amplitude. Many orange/red patterns are observed in the plots that represent trajectories crossing some synchro-betatron resonance lines.

The CM analysis in the $x$-$y$ plane (left panel of Fig. 4(a)) also exhibits distinct patterns. The red patterns in Fig. 4(a) illustrate the stable boundary that agree well with the boundary of the DA in the $x$-$y$ space. Some trajectories having large residual convergence errors relative to a pure rotation are highlighted in Fig. 4, and their locations are found to agree well with the resonance structures observed in the FMA results in Fig. 4(b). Using only the one-turn transfer map, the CM method therefore effectively probes the stable region, demonstrating both its efficiency and accuracy for DA evaluation.

\begin{figure}[!htb]
\centering
\includegraphics[width=1\columnwidth]{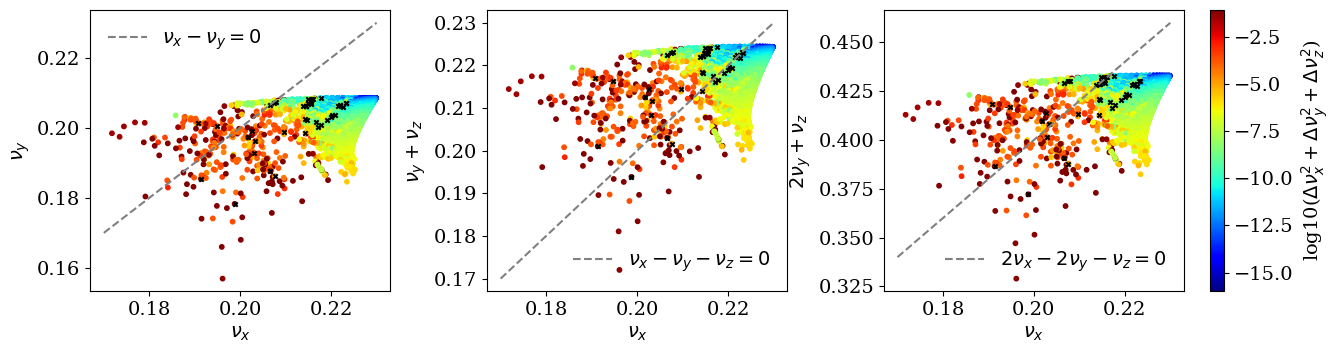}
\caption{FMA results in frequency space. The left plot shows the result in $\nu_x$ and $\nu_y$ space, the middle one shows the $\nu_x$ and $\nu_y+\nu_z$ space, and the right one shows the $\nu_x$ and $2\nu_y+\nu_z$ space. The black cross in the plots correspond to the selected points in Fig. 4.}\label{fig5}
\end{figure}

Figure 5 shows the trajectories highlighted in Fig. 4 in frequency space. It is found that these selected points lie on three different resonance lines, including the linear transverse resonance $\nu_x=\nu_y$, and synchro-betatron resonances such as $\nu_x=\nu_y+\nu_z$ and $2\nu_x-2\nu_y-\nu_z=0$. These resonances strongly affect long-term beam stability and can lead to particle loss, indicating highly nonlinear dynamics in the ring.

Figure 6 provides a direct comparison between the CM residual and DA boundaries obtained from tracking over $10^3-10^6$ turns. The transition from strongly convergent to poorly convergent CM solutions follows the progressive reduction of the tracked stable region, although the CM boundary does not coincide exactly with the $10^6$-turn DA. This comparison illustrates the intended role of CM as a fast indicator rather than a replacement for long-term DA tracking.

\begin{figure}[!htb]
\centering
\includegraphics[width=0.6\columnwidth]{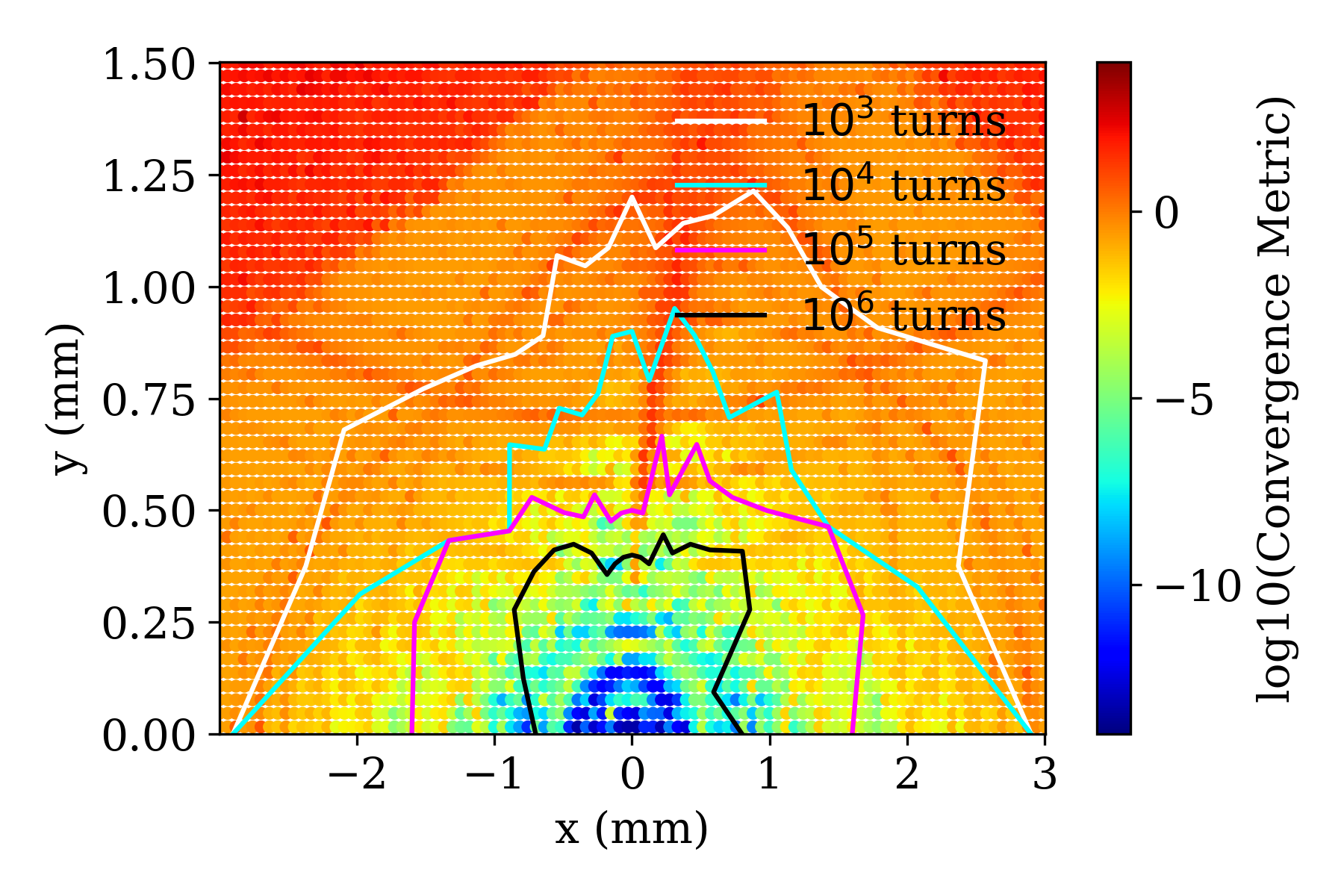}
\caption{Comparison of the DA boundary of the HSR and the CM results. The white, cyan, pink and black lines represent tracked DA boundaries over $10^3$, $10^4$, $10^5$, and $10^6$ turns, respectively. }\label{fig6}
\end{figure}

\begin{figure}[!htb]
\centering
\includegraphics[width=1\columnwidth]{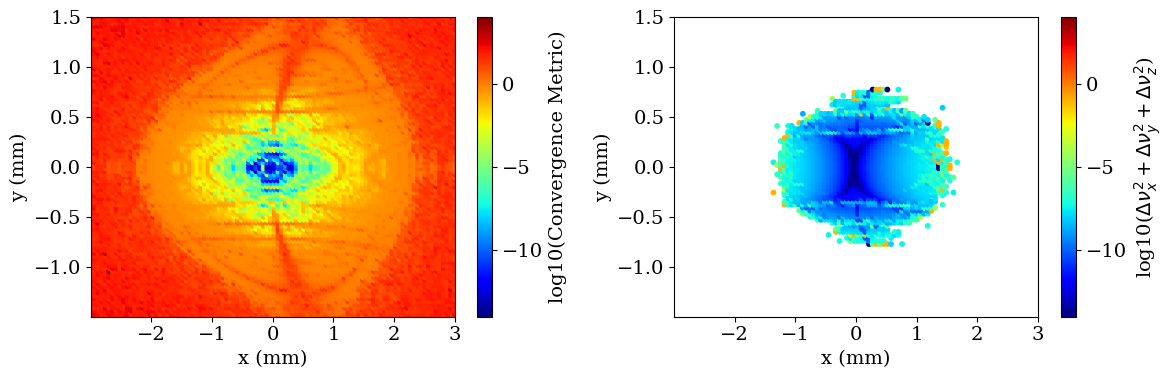}
\caption{Results of applying the CM method (left) and FMA (right) to the HSR with increased multipole components in the crab cavities. The strength of sextupole components in the crab cavity is increased to $10^{-2}$ m$^{-3}$.}\label{fig7}
\end{figure}

To further study the impact of multipole components in crab cavities, we increase the sextupole strength from $10^{-3}$ m$^{-3}$ to $10^{-2}$ m$^{-3}$. With the increased multipole components, both CM and FMA show a significant reduction in the DA size by nearly a factor of two in Fig. 7. This observation is consistent with previous findings \cite{barranco2016long, wu2022eic}, revealing the sensitivity of beam dynamics to nonlinearities introduced by multipole components in crab cavities. Since the multipole components of the crab cavities are time dependent, traditional magnets cannot compensate for their nonlinear effects. Therefore, it is important to control the multipole strength during crab cavity design or use external magnets to cancel these effects to maintain sufficient DA.

\section{Conclusion}
The square matrix-based CM is extended to the full 6-D phase space to study nonlinear beam dynamics in ring accelerators. By reformulating the one-turn map to a square matrix representation through eigen-decomposition, a set of approximated action-angle variables are constructed. The residual convergence error from transforming the map into a pure rotation is used as an indicator of trajectory stability.

We demonstrate that in a toy model with crabbing kicks, the 6-D CM reproduces the main resonance structures observed in the corresponding FMA analysis and highlights additional structures associated with the designed resonance condition. This method is then applied to the the EIC HSR lattice, where it effectively identifies resonance structures that critically impact long-term beam stability. By increasing the strength of the multipole component in the crab cavities, a significant reduction in DA is observed. The CM identifies major resonance patterns and stability-boundary-related structures that are qualitatively consistent with the FMA results. Compared to particle tracking simulation that can require more than $10^6$ turns of tracking, the CM method allows efficient evaluation of long-term beam stability with the one-turn transfer map, and can serve as a fast indicator for DA. 

This work establishes the 6-D CM method as a fast and useful complementary method for nonlinear beam dynamics analysis. Its ability to detect high-order resonances and high efficiency from using only one-turn map makes it a fast indicator for future accelerator design, optimization, and long-term stability studies.

\begin{acknowledgments}
This work is supported by the DOE Office of Science, with award number DE-SC0024170.
\end{acknowledgments}

\end{document}